# Sponge like nanoporous single crystals of gold

**Maria Koifman Khristosov[1,2], Leonid Bloch[1], Manfred Burghammer[3], Yaron Kauffmann[1], Alex Katsman[1] and Boaz Pokroy[1,2,\*]**

[1]Department of Materials Science and Engineering, Technion − Israel Institute of Technology, 32000 Haifa, Israel

[2]Russell Berrie Nanotechnology Institute, Technion − Israel Institute of Technology, 32000 Haifa, Israel

[3]European Synchrotron Radiation Facility, BP 220, F-38043 Grenoble Cedex, France, and X-ray Microspectroscopy and Imaging Research Group, Department of Analytical Chemistry Ghent University, B-9000 Ghent, Belgium

\*Correspondence should be addressed to B.P. (bpokroy@technion.ac.il).

**ABSTRACT**

Single crystals in nature often demonstrate fascinating intricate porous morphologies rather than classical faceted surfaces. We attempt to grow such crystals, drawing inspiration from biogenic porous single crystals. Here we show that nanoporous single crystals of gold can be grown with no need for any elaborate fabrication steps. These crystals are found to grow following solidification of a eutectic composition melt that forms as a result of the dewetting of nanometric thin films. We also present a kinetic model that shows how this nano-porous single-crystalline structure can be obtained, and which allows the potential size of the porous single crystal to be predicted. Retaining their single-crystalline nature is due to the fact that the full crystallization process is faster than the average period between two subsequent nucleation events. Our findings clearly demonstrate that it is possible to form singe crystalline nano porous metal crystals in a controlled manner.

**INTRODUCTION**

Crystals grown in the laboratory by classical methods of nucleation and growth have facets dictated by their atomic structure and by minimization of the surface free energy of the different crystallographic planes[1]. These features result in faceted crystals in which the revealed planes are the low-energy ones. Remarkably, however, biogenic crystals produced by living organisms often demonstrate unfaceted single crystals with rounded, intricate, and even porous structures. Such single crystals can be produced via an amorphous precursor phase in which the precursors can be molded into any desired shape prior to their crystallization[2-4].

The achievement of intricately shaped nanoporous single-crystalline materials is undoubtedly of considerable scientific and technological potential[5]. It was previously shown that micron-sized porous single crystals of calcite can be grown via bio-inspired routes[6,7].

Even porous metal materials (PMMs) that are polycrystalline are functionally promising[8,9] for research and industry in the fields of chemistry[10], medicine[11], environmental sciences, energy storage and others[12]. Besides the intrinsic properties of monolithic metals, in PMMs the new and attractive properties derived from the shape, size and distribution of pores include changes in specific surface area, permeability and capillarity to mention just a few. Owing to their unique structure, such porous metals are widely employed in catalytic support[13], filtration[14], separation[15], heat exchange[16], fuel cells[17], and many additional uses. Common methods of preparing nanoporous metals are dealloying**[18]**, or depositing the metal within a 3D mold[19] such as a latex sphere[20] or other soft templates[21].

A particularly important example of a PMM is nanoporous gold, which has great potential for applications in catalysis[22,23], sensors[24], actuators[25], and electrodes for electrochemical uses[26]. The yield strength of nanoporous gold is very high, equal to or higher than that of bulk Au, while the nanoporous material also has a much lower density and very large surface area[27]. A common way to obtain nanoporous gold is by dealloying of an Au-Ag alloy via selective removal of Ag in a corrosive environment such as a nitric acid solution[28].

The nanoporous gold prepared by dealloying has a nanocrystalline structure[29]. Formation of nanoporous gold from an original Au-Ag micron-sized grained ingot has been reported[30], but it was prepared by dealloying, which resulted in a nanocrystalline microstructure. This result can be clearly observed from the decrease in X-ray coherence length, observed in the broadening of the diffraction peaks and the increase in lattice misorientations. A few examples of single-crystalline porous gold have been described, but these nanometric structures were only two-dimensional[31]. Besides possessing the attractive properties of polycrystalline porous gold, a nanoporous *single crystal* of gold is likely to have additional novel properties. Studies have shown that grain boundaries in thin films of gold enhance the electrical resistivity of nanometric thick films owing to grain boundary-induced scattering of the electrons[32-34]. Not only would the conductivity of the porous single crystal be higher than that of porous polycrystalline gold, but as single crystals their 3D crystallographic orientation could also be selected. The higher conductivity expected from such nanoporous structures could increase the efficiency of electrodes for electrochemical supercapacitors, as described elsewhere[26]. Another advantage of the absence of grain boundaries in such structures is to be found in the higher thermal stability of their micro- and nano-framework. This results from lower rates of self-diffusion, which are attributable in turn to the total elimination of grain boundary diffusion routes. The latter adventitious property could be helpful in cases where such structures are utilized in catalysis when operational temperatures in the range of 20−200°C are needed[35,36]. As one of the most prevalent applications of nanoporous gold is in catalysis, we feel that this property could be interesting for other researches in this field.

In this paper we present an unprecedented method, using simple steps, to grow micron-sized nanoporous single crystals of gold with intricate morphologies. Unlike any other examples to date, our structures are formed via thermodynamically driven self-formation. Thin films of two components are evaporated onto a nonreactive surface, followed by heating to above the eutectic temperature. This procedure causes the films to melt, producing a eutectic melt on the surface. Because of the low surface energy of the appropriately chosen substrate, dewetting takes place and isolated droplets of the melt are produced all over the substrate's surface. Fast cooling preserves the shapes of both the droplets and the eutectic microstructure. Finally, by selectively etching of one of the components, a porous single crystal structure of the other component is obtained. We demonstrate this general principle on an Au-Ge eutectic system, which yields nanoporous single crystals of gold. Furthermore, we present a model that explains our experimental findings and enables us to calculate the conditions needed to obtain rationally designed single-crystalline porous structures. It also enables us to predict the size limits of such an approach.

## RESULTS

**Preparation process description**

Thin films comprising two layers were evaporated at relative thicknesses appropriate for the eutectic ratio, which in the Au-Ge system is 28at% Ge[37]. Annealing above the eutectic temperature resulted in melting of these thin films and formation of a eutectic melt. Choice of $SiO_2$ as a substrate enabled droplets of this melt to be formed by dewetting. Cooling to below the eutectic temperature resulted in solidification of the eutectic liquid to form a eutectic-like solid structure. Wet etching of the samples to remove the germanium revealed droplets of porous gold, as seen in **Fig. 1a-b** by high resolution scanning electron microscopy (HRSEM). Analysis of the microstructure by energy-dispersive X-ray spectroscopy (EDS) confirmed that the Ge was fully removed during the etching process (Supplementary Table 1). Using a focused ion beam (FIB) we were able to visualize a cross section of the droplets and obtain a view of the gold microstructure within a droplet (**Fig. 1c-d**).

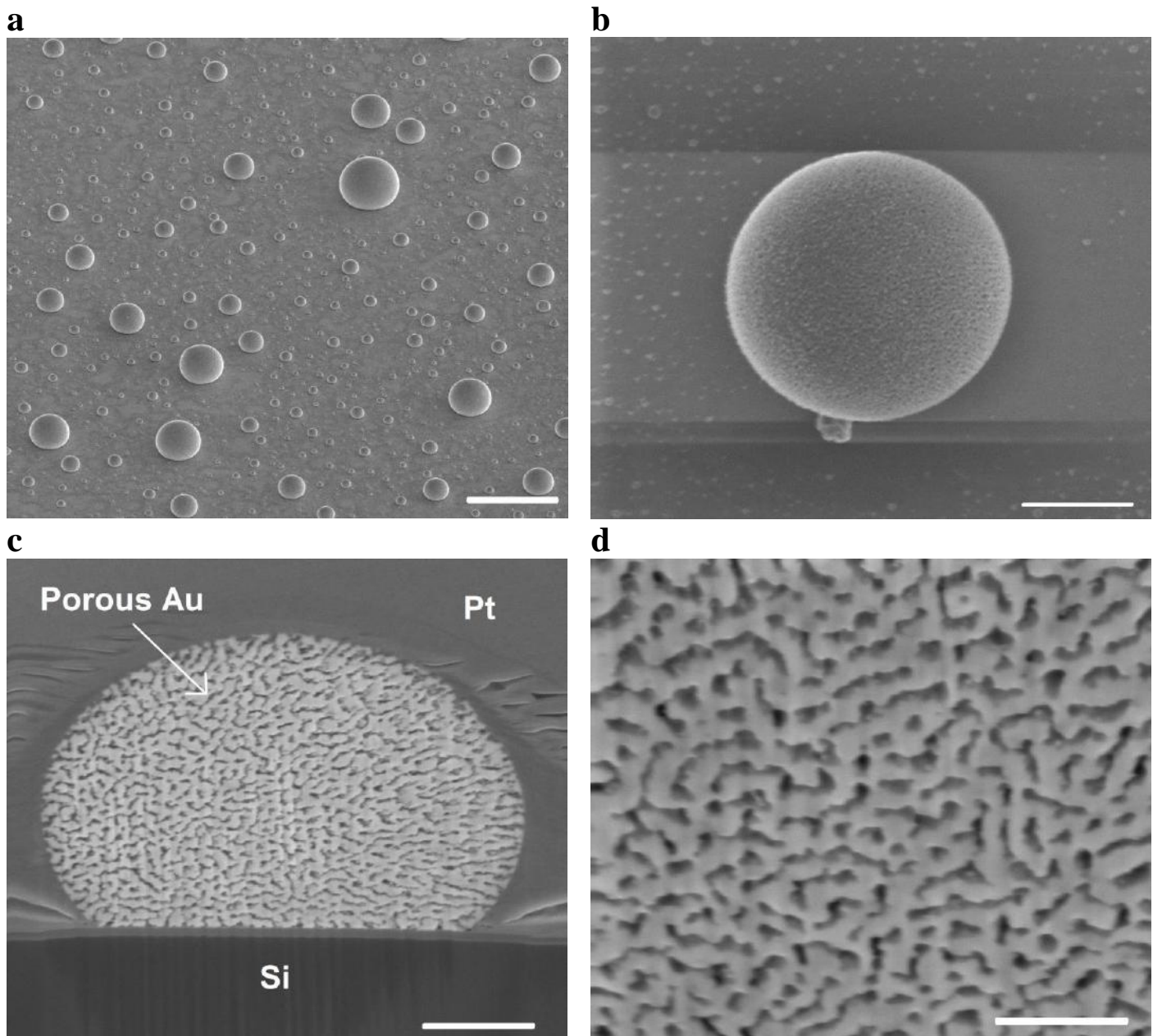

**Figure 1 Gold nanoporous single crystal particles.** (**a**) High resolution scanning electron microscopy (HRSEM) micrograph of the droplets on a SiO$_2$ surface, large area view (scale bar, 20 μm). (**b**) HRSEM micrograph of a eutectic gold droplet after wet etching, top view (scale bar, 2 μm). (**c**) HRSEM micrograph of a cross section of the droplet shown in **b**, obtained using a focused ion beam (FIB), reveals the nano-porous structure of gold after wet etching of Ge (scale bar, 1 μm). (**d**) HRSEM micrograph of high magnification in (**c**) (scale bar, 500nm).

**TEM diffraction of the gold nanoporous single crystal**

A thin cross section of a sample droplet was prepared by FIB and examined by transmission electron microscopy (TEM). A high-angle annular dark-field scanning TEM (HAADF-STEM) micrograph of the cross section is presented in **Fig. 2a**. Selected area diffraction on an area with a diameter of approximately 4 μm in the center (the main part) of the droplet (**Fig. 2b**) revealed a diffraction pattern characteristic of a single crystal that could be fully indexed within the gold structure (**Fig. 2c**). This indicates that the gold in the eutectic-like structure within the droplet had solidified into an intricately shaped single crystal.

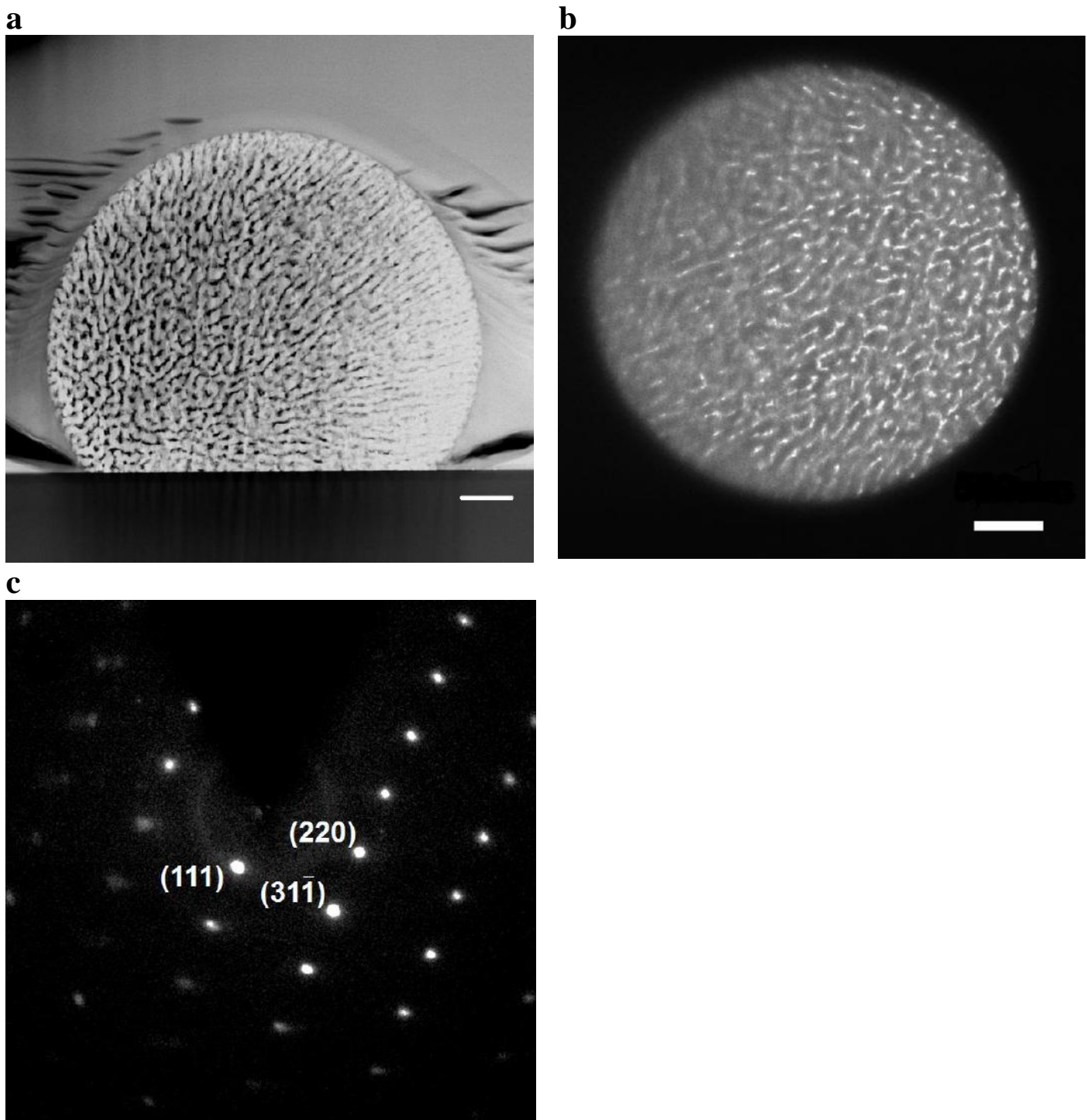

**Figure 2 Transmission electron microscopy (TEM) investigation of the nanoporous single crystal. (a) A high-angle annular dark-field scanning TEM** (HAADF-STEM) micrograph of a cross section of a droplet, showing microstructure of gold after Ge etching

(scale bar, 500 nm). (**b**) TEM image of the area of the droplet from which the diffraction pattern was acquired, showing the selected area aperture in **a** (scale bar, 500 nm). (**c**) Diffraction pattern taken from the area shown in **b**, fully indexed within the gold structure; ZA (zone axis) [121].

**Synchrotron scanning diffractometry**

To further verify that this porous gold is indeed a single crystal we utilized another state-of-the-art characterization technique, namely synchrotron submicron scanning diffractometry (ID13; ESRF, Grenoble, France), on a FIB-sectioned gold crystal. The same cross section as that of the droplet investigated by TEM (**Fig. 2**) was now examined by scanning diffraction. Diffraction patterns were scanned over the entire cross section of the droplet (**Fig. 3**). **Fig. 3a** depicts a map of a single reflection ({200} plane, 0° rotation). The same reflection occurs throughout the droplet at nearly the same radial and azimuthal coordinates, with only some slight variations in intensity. **Fig. 3b** depicts the average of all the diffraction scans taken from the entire area of the droplet. This can be compared with **Fig. 3c**, which shows a single diffraction from the droplet's center. From a comparison of these images it is evident that the major reflections are identical, and do not shift their positions even though the figure integrates a large number of individual scans (317 in all). **Fig. 3d** represents the azimuthally regrouped central diffraction with markings for gold crystallographic planes, and **Fig. 3e** depicts a rocking curve for the same reflection. Shown are the mean intensity over the droplet and its 10%, 30% 70% and 90% percentiles. Results for the {220} planes are shown in (Supplementary Fig. 1). The results obtained from the TEM and the scanning diffractometry provide conclusive proof that the nanoporous gold crystals are indeed single crystals.

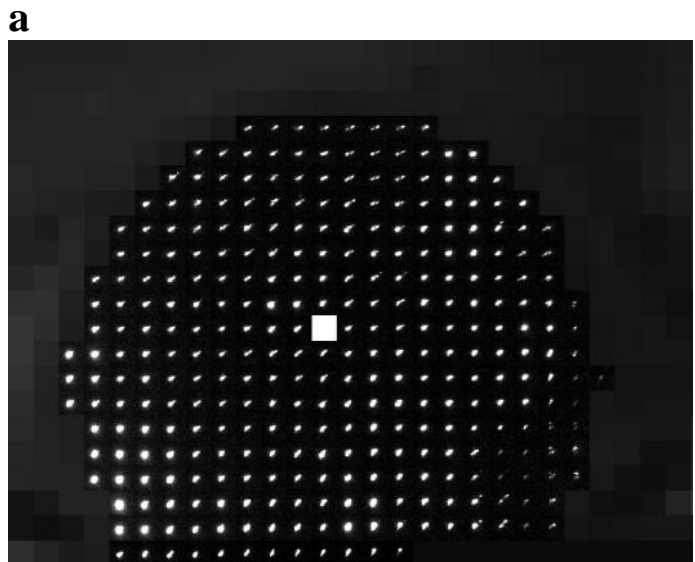

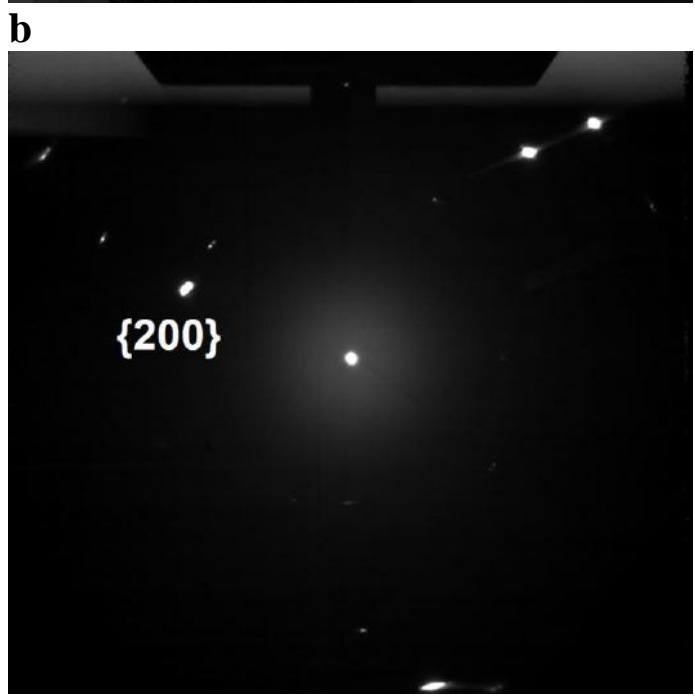 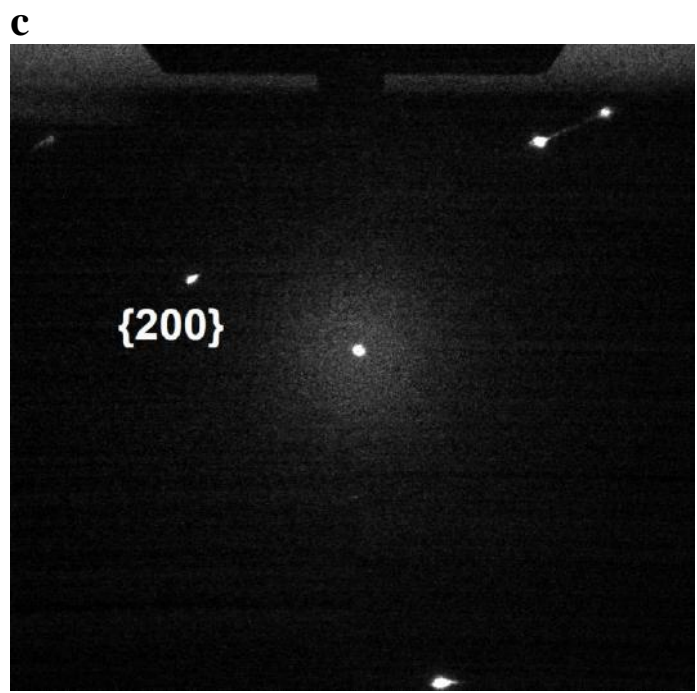

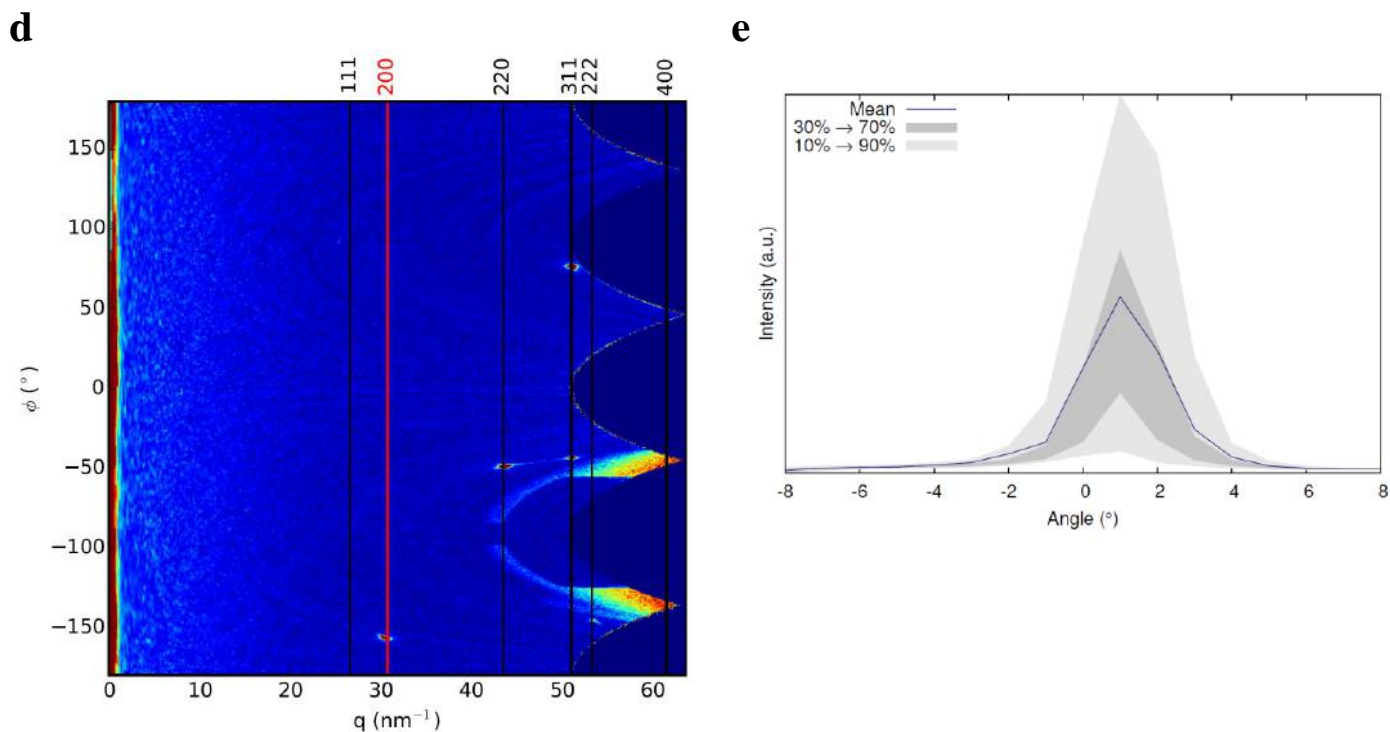

**Figure 3 Synchrotron submicron scanning diffractometry of the gold nanoporous single crystal.** (**a**) Reflection map for the {200} planes of the sample shown in Figure 2. (**b**) The average diffraction pattern from the drop area. (**c**) Diffraction pattern from the center of the drop, marked with a white square on a. (**d**) Azimuthally regrouped central diffraction pattern, $q = 2\pi/d$, $\varphi$ – radial axis. (**e**) Rocking curve of a reflection from the {200} planes.
10

**Controlling the micro and nanostructures**

By altering the cooling rates of the samples, micro- and nanostructures of different sizes could be obtained (**Fig. 4**). Ligament sizes in the solid droplet after cooling at a fast rate (35 °C s$^{-1}$) were about 57 ± 12 nm for gold and 43 ± 8 nm for germanium (**Fig. 4a**). At a slow cooling rate (0.6 °C s$^{-1}$), the approximate ligament sizes of the gold and the germanium were 300 nm (or larger) and 39 ± 8 nm, respectively (**Fig. 4b**). This phenomenon resulted from a coarsening of the solid eutectic phase during the longer cooling phase.

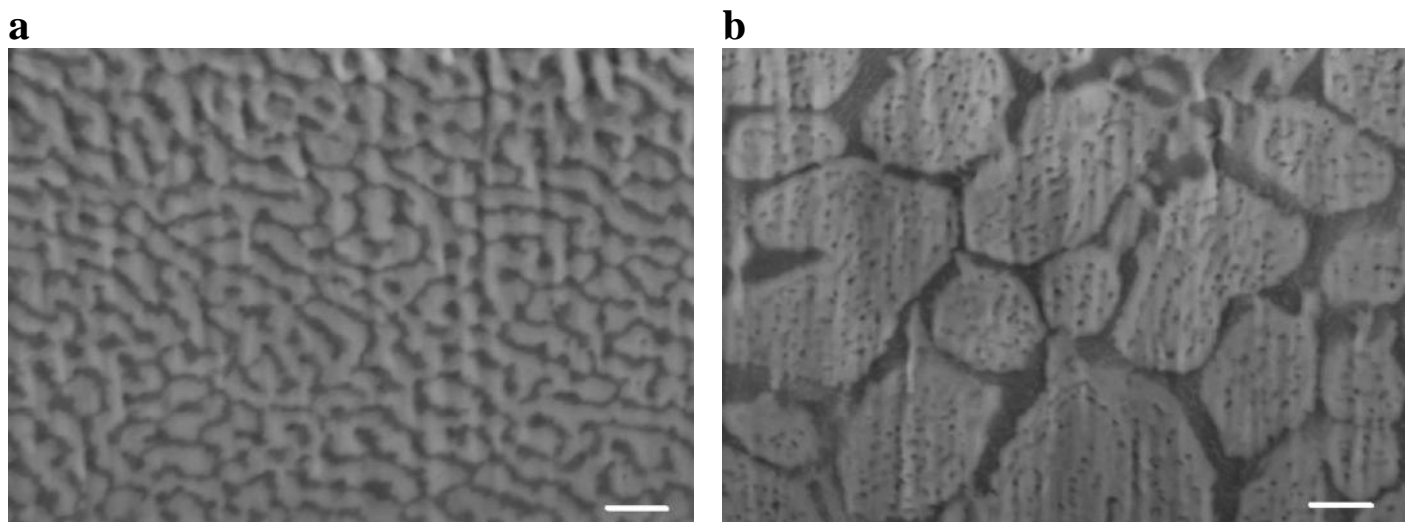

**Figure 4 Controlling the micro and nanostructures of the single crystal nanoporous gold.** HRSEM micrograph of a FIB cross section of the eutectic microstructure at (**a**) 35 °C s$^{-1}$ and (**b**) 0.6 °C s$^{-1}$. Scale bar, 200 nm.

**Surface area and thermal stability**

In addition, by analyzing the cross-sectional HRSEM micrographs we estimated the surface area of the porous Au crystals. For this purpose we assumed a homogeneous microstructure and used a previously described methodology[38]. Our calculations yielded an estimated surface area of 3.1 m$^2$g$^{-1}$, which is in the range of other known nanoporous metals[39-41] approximately 3 to 10 m$^2$g$^{-1}$.

We then compared the thermal stability of our single crystalline nanoporous nanostructures to that of nanoporous gold prepared by the dealloying method[28]. The results are shown in Supplementary Fig. 2-5. After annealing both types of porous gold samples, one prepared by our eutectic decomposition method as described here, and the other by dealloying, at 200°C for 15 min and for 45 min, and at 250°C for 15 min, we compared their nanostructures. The samples prepared by the dealloying method failed at lower treatment time and tempertaures: grain boundaries developed into cracks, porous particles were found to be damaged at the edges and coarsening was clearly observed. In conspicuous contrast, samples prepared by our method retained their porous nanostructure and showed no evidence of damage.

## DISCUSSION

By using state-of-the-art characterization techniques such as TEM and synchrotron submicron scanning diffractometry (ID13), we were able to show that the porous gold structure within eutectic droplets is a single crystal. Use of these techniques further revealed continuous growth of a single gold crystal during formation of the eutectic-like two-phase structure during its solidification. Although nanoparticles of nanoporous gold have been demonstrated previously [19-21,28], our method is novel in that we do not use the dealloying mechanism (a "top-down" approach); instead, our porous single crystals grow from a liquid phase during eutectic decomposition, which is a spontaneous "bottom-up" approach. Growth of the porous single crystal by the latter method occurs rapidly and is a *self-forming* process (eutectic decomposition). The outcome is a nanoporous single crystal of gold, with the germanium residing only in the pores. Another advantage of our method is that by means of a simple, fast, and low-cost process we obtain free-standing porous particles whose shape can be determined by the original droplets wetting properties and which can easily be used in other processes with no need to cut the bulk material by any type of fabrication technique.

When comparing our method with the conventional dealloying procedure in the formation of nanoporous gold, it should be borne in mind that not all that appears to be a single crystal is indeed a single crystal. Several studies[30,42] have implied that given the apparently unchanged morphology and orientation of the original grains before and after dealloying, the nanoporous gold products are still single crystals of the same original size. This was demonstrated on the basis of grain morphology and electron back-scattered diffraction in SEM. The only way to unambiguously prove that a grain is in fact a single crystal is by the use of high-resolution X-ray diffraction techniques, as we did in this study. For further verification, we performed in-situ dealloying with high-resolution powder diffraction at the synchrotron (ID22; ESRF) on $AuAl_2$ intermetallic. The results clearly showed how the coherence length of the gold that remains after dealloying is drastically reduced (Supplementary Fig. 6). Though the structure of $AuAl_2$ is different than that of pure gold (both are cubic), the literature shows the exact same phenomenon for deploying silver from gold[30].

One of nano porous gold promising applications is a low temperature catalytic CO oxidation[35,36] to $CO_2$. This process can be undertaken at temperatures in the range of 20-200°C, and usually is performed utilizing nanoporous gold prepared by dealloying or gold nanoparticles. The disadvantage of the nanoporous gold prepared by dealloying is coarsening of the nanoporous structure due to the temperature-induced diffusion. Although our samples have slightly larger ligament size as compared to the ligament size of the dealloyed gold counterpart, and therefore slightly lower surface area, the latter demonstrate superior thermal stability. This relative improvement is due to the lack of grain boundaries within single crystals.

Discerning the mechanism governing the growth of nanoporous crystals that maintain the characteristics of a single crystal is important not only for basic scientific knowledge but also to allow this technique to be applied, in a controlled manner, in technologically important crystalline systems. We therefore felt it necessary to develop a kinetic model that could explain how these nanoporous single crystals are formed, and would allow us to estimate the maximum achievable size of such crystals.

In the case of a fully eutectic structure, the volume ratio $V_{Au}/V_{Ge} = 1.92$ (calculated on the basis of the eutectic composition) enables gold crystals to grow continuously throughout the entire droplet, accompanied by repeated nucleation and growth of rod-like germanium crystal channels surrounded by a gold matrix. When considering a crystallization process in the Au-Ge eutectic system it should be remembered that normal crystallization of a non-equilibrium eutectic liquid usually starts with the heterogeneous nucleation of one solid phase, in the present case probably Au. The nucleation also most probably takes place at the liquid/substrate interface. In the next stage the adjacent liquid is substantially enriched by Ge, providing heterogeneous and, if possible, epitaxial nucleation of Ge crystals at the surface of this Au crystal. The formation is governed by the free energy gain as a result of undercooling to below the eutectic temperature, $\Delta T = (T_{eut} - T)$, while atoms are being transferred from the liquid phase to the growing crystal:

$$\Delta G_{tr} = (T_{eut} - T)\Delta S_{tr}, \tag{1}$$

where $\Delta S_{tr}$ is the entropy change in the L (eutectic) $\to$ Au(s) + Ge(s) transformation. Obviously, additional energy is needed to create the solid/liquid and Au(s)/Ge(s) interfaces. The epitaxial nucleation and growth of Ge on Au is possible in the crystallographic orientation relationships $Au(111)[11\bar{2}]//Ge(111)[11\bar{2}]$ and $Au(\bar{1}\bar{1}1)[11\bar{2}]//Ge(111)[11\bar{2}]$ [43], which may provide relatively low interfacial energies. Growth of the eutectic structure is limited by volume diffusion in the liquid in front of the freezing solid/liquid interface. The growth rate can be evaluated according to the simple model suggested by Turnbull[44]:

$$V = kD\frac{\Delta X}{\lambda} = kD\frac{\Delta X_0}{\lambda}\left(1 - \frac{\lambda^*}{\lambda}\right), \tag{2}$$

where $D$ is the diffusion coefficient in the liquid, $\Delta X$ is the difference in composition of the liquid above the freezing interface across the eutectic lamellar spacing $\lambda$, $\Delta X_0$ is the difference in composition for $\lambda \to \infty$, k ~ 1 is a geometry-dependent coefficient, and $\lambda^*$ is the critical lamellar spacing defined by Zener[45]:

$$\lambda^* = \frac{2\gamma_{\alpha\beta} v_{mol}}{\Delta G_{tr}} = \frac{2\gamma_{\alpha\beta} v_{mol}}{\Delta T \Delta S_{tr}}, \qquad (3)$$

where $\gamma_{\alpha\beta}$ is the energy of the solid/solid (Au/Ge) interface and $v_{mol}$ is the liquid molar volume. Using the transformation entropy value for the Au-Ge system[46] $\Delta S_{tr}$ = 23.9 Jmol$^{-1}$·K$^{-1}$, and assuming $\gamma_{\alpha\beta}$ = (0.2÷0.4) Jm$^{-2}$ and $\Delta T$ = (10÷20)$K$, we can estimate from eq. (3) the critical width of lamellae as $\lambda^*$ = (8÷32) nm, which is comparable to the half of the observed ligament size (40 ÷ 60) nm. The relatively minor observed supercooling is attributable to the heterogeneous character of the nucleation. The supercooling corresponds to a difference in composition of the couple zone of the Au/Ge phase diagram[46] $\Delta X_0$ = 0.01 ÷ 0.02. From viscosity experiments at high temperatures[47] the self-diffusion coefficient in the liquid gold was found to be in the range of 2.02·10$^{-9}$ to 3.35·10$^{-9}$ m$^2$s$^{-1}$ at temperatures between 1063°C and 1364°C with an activation energy of $E_a$ = 0.316 eV. The diffusion coefficient in the undercooled eutectic liquid can be evaluated by extrapolation of these values to temperatures below the Au\Ge eutectic temperature. $D = D_0 \exp(-E_a/k_B T)$ with $D_0$ = 3.1·10$^{-8}$m$^2$s$^{-1}$, yielding ~8.5·10$^{-11}$m$^2$s$^{-1}$ for the eutectic temperature. We can now estimate the growth rate and the time to achieve full crystallization of the eutectic structure. The growth rate is evaluated as $V$ = (10 ÷ 30) μms$^{-1}$ and the total crystallization time in droplets (2 ÷ 5) μm in size is $\tau_c = R_d/V$ = (0.07 ÷ 0.5)s, where $R_d$ is the droplet size.

A more sophisticated theory of steady-state eutectic growth was developed by Jackson and Hunt (JH)[48] and then modified by several authors[49-51]. The estimations based on this theory are presented in Supplementary Note 1. According to evaluations based on the JH theory, the freezing rates, $V = (30 \div 60)$ µms$^{-1}$, are slightly higher (and the total crystallization times are lower) than those obtained from the Turnbull model[44].

The rate of heterogeneous nucleation of gold crystals, presumably at the substrate/liquid interface, can be deduced on the basis of classical kinetic theory[52,53]:

$$J_{ss} = J_0 \exp(-W^*/kT), \qquad (4)$$

where $W^*$ is the height of the heterogeneous nucleation barrier and $J_0$ is a pre-exponential factor. As shown in Supplementary Note 2, for reasonable material parameters the pre-exponential factor can be evaluated as $J_0 \approx (4 \div 6) \cdot 10^{20}$ s$^{-1}$µm$^{-3}$, and at the moment when the first stable nucleus appears, the heterogeneous nucleation rate in the droplet reaches $J_{ss}(t_1) = J_0 \Delta V_d e^{-y_1^2} \approx \frac{2y_1^3 \alpha}{B'}$ nuclei per second (see Supplementary Note 2), which for the cooling rate $\alpha = 1$ C°s$^{-1}$ (used in most of the present experiments) provides about $(0.8 \div 1.0)$ nuclei per second. Comparison of the average time $\Delta t_{12} \approx \frac{B'}{\alpha} \frac{\ln 2}{2y_1^3}$ between the two first nucleation events with the time of full crystallization of the droplet $\tau_c$ allows us to formulate a criterion for the appearance of a monocrystal porous structure described in the previous sections:

$$\chi = \frac{\Delta t_{12}}{\tau_c} = \frac{VB'}{R_d \alpha} \frac{\ln 2}{2y_1^3} \geq 1. \qquad (5)$$

For typical values of parameters (see Table 2 in Supplementary Note 2): $B' = 700$ °C, $y_1 = 6.5$, $\chi \approx \frac{0.88V}{R_d \alpha (°C s^{-1})}$. For droplets with radii $R_d \leq 10$ µm and cooling rates $\leq 1$°C s$^{-1}$, with $V = (30 \div 50)$ µms$^{-1}$, formation of Au-Ge eutectics with the monocrystal gold matrix seems to be highly probable. On the other hand, with faster cooling, for example with the rate $\alpha = 35$ °C s$^{-1}$, such a structure is unlikely to be formed. In other words, for cooling rates ~1 °C s$^{-1}$ the second nucleus appears $(0.7 \div 0.9)$ s on average after the first one, and this period is long enough to allow full crystallization of the (2 ÷ 10) μm droplets before the second nucleation event occurs. At the same time, for droplets with radii R ≥ 20 μm and/or high cooling rates $\alpha \geq 10$ °C s$^{-1}$, several nuclei can appear during crystallization.

In conclusion, in each droplet during cooling of the sample, two processes are competing – nucleation and crystal growth. The times taken for the full crystallization process in the micron-sized droplets investigated here appeared to be substantially shorter than the average period between the two subsequent events of nucleation of gold crystals from the eutectic melt. This can explain why the gold matrix of the eutectic structure was formed as a single crystal. To further validate our kinetic model we studied how changing the cooling rates of the process affects the number of single crystal domains in each droplet. As our model suggests, an increase in the cooling rate from 0.6°C s$^{-1}$ to 35°Cs$^{-1}$ is followed by a clear transition from single crystal to polycrystalline droplets (Supplementary Fig. 7). To conclude, using the relatively simple method of dewetting from thin films, and by exploiting thermodynamic phenomena such as crystallization from a eutectic melt, we demonstrated here that it is possible to produce nanoporous gold single crystals. We have developed a kinetic model that explains this phenomenon and shows that the full crystallization process is faster than the average period between two subsequent nucleation events, a key factor allowing the intricate 3D single crystals of gold to retain their single-crystalline nature. Based on our calculations, we can predict that this method allows for the growth of nanoporous single crystals of gold of up to several hundred microns in size. We also clearly showed that nanoporous single crystal prepared by eutectic composition demonstrate superior thermal stability as comapred to their counterpart nanoporous gold prepared by dealloying, which is essential for catalysis. We believe that this method can also be employed in other crystal systems, thereby opening the door to new technological capabilities.

## METHODS

### Sample preparation

SiO$_2$ (100 nm) was grown on (001) Si wafers by thermal oxidation at 1100 °C. The oxide layer served as a diffusion barrier for preventing Si migration from the wafer. Gold and germanium films (99.999% pure, Sigma-Aldrich) were successively evaporated onto the SiO$_2$ substrate in an e-beam-equipped AircoTemescal FC-1800 evaporating system under a high vacuum of 10$^{-7}$ Torr at room temperature, yielding a deposition rate of 8 Ås$^{-1}$. The gold and germanium films were 150 nm and 78 nm thick, respectively. Samples were thermally annealed at 550°C for 5 min in a MILA-5000 ULVAC-RIKO rapid thermal annealer or in a Jiplec JetFirst-100 rapid thermal annealer in an ambient flow of Ar-H$_2$ (15% H$_2$) (99.999%, 150 sccm) or vacuum (10$^{-5}$ Torr) at a heating rate of 10 °C s$^{-1}$. The cooling rates were 35 °C s$^{-1}$ and 0.6 °C s$^{-1}$. Samples were wet-etched in two steps: (i) immersion in a solution of NH$_4$OH:H$_2$O$_2$ (1:25% vol) for 1h and then rinsed, first in deionized water and then in ethanol. (ii) immersion in KOH solution (1.25M) for 16h and then rinsed again in water and ethanol. Cross-sectional samples were prepared by an FEI Strata 400S dual-beam focused ion beam (FIB). Low-voltage argon ion milling was then applied for final thinning and cleaning of surface using the Gentle Mill, model IV8 (Technoorg Linda). Thermal stability experiments were performed in Jiplec JetFirst-100 rapid thermal annealer in 200-250°C for altering times in vacuum (10$^{-5}$ Torr).

### Sample characterization

Surfaces were imaged with a Zeiss Ultra Plus high-resolution scanning electron microscope (HRSEM), combined with energy-dispersive X-ray spectroscopy (EDS). Images of the cross sections were acquired with the Strata 400S dual-beam FIB. Transmission electron microscopy (TEM) imaging and electron diffractions were obtained using an FEI C$_s$ corrected Titan 80-300 KeV FEG-S/TEM operated at 300 KeV. Single-crystalline droplets were characterized by nanofocus X-ray beam analysis on ID13 of the European Synchrotron Research Facility (ESRF), Grenoble, France. The samples were rotated from −45 to 45 degrees, with 1-degree intervals. After each interval the sample was scanned with an X-ray beam at a wavelength of 0.83189Å, focused to approximately 200×150 nm (full width at half maximum) at the sample plane.

## ACKNOWLEDGEMENTS

Thin films were fabricated at the Micro-Nano Fabrication Unit (MNFU) at the Technion − Israel Institute of Technology, Haifa. We are grateful to Dr. Tzipi Cohen-Hyams, Dr. Alex Berner, and Michael Kalina for their help in preparing samples and operating the electron microscopes. We thank Boris Haimov for useful discussions, and Avigail Aronhime Bracha and Yael Etinger for helping with sample preparation. We thank Prof. Eugen Rabkin for use of the rapid thermal annealer. The research leading to these results received funding from the European Research Council under the European Union's Seventh Framework Program (FP/2007–2013)/ERC Grant Agreement no. 336077. M.K.K. is grateful for financial support by the Israeli Ministry of Science, Technology and Space. Nanofocusing lenses were kindly provided by the group of Prof. Dr. Christian Schroer from TU Dresden, Germany.

**Supplementary Information** is available in the online version of the paper.

## AUTHOR CONTRIBUTIONS

B.P. conceived the work; M.K.K performed the experiments; L.B. analyzed the synchrotron data; A.K. developed the theoretical model; M.B. performed the synchrotron measurements; Y.K. performed the TEM measurements; M.K.K. and B.P. wrote the manuscript with comments from all authors.

## COMPETING FINANCIAL INTERESTS

The authors declare no competing financial interests.